\documentclass[a4paper,11pt]{article}
\usepackage{plain}

\usepackage{pos}

\title{The standard model in Wilson lattice gauge theory}
 \ShortTitle{Standard model on the lattice}

\author*[a]{Michael Creutz}

\affiliation[a]{Brookhaven National Laboratory,\\
  Upton, NY 11973, USA}

\emailAdd{mike@latticeguy.net}

\abstract{The $SU(3)\otimes SU(2) \otimes U(1)$ standard model maps
  smoothly onto a conventional Wilson lattice gauge formalism,
  including the parity violation of the weak interactions. The
  formulation makes use of the pseudo-reality of the weak group and
  requires the inclusion a full generation of both leptons and
  quarks. As in continuum discussions, chiral eigenstates of the Dirac
  operator generate known anomalies, although with rough gauge
  configurations these are no longer exact zero modes.}

\FullConference{The XVIth Quark Confinement and the Hadron Spectrum Conference (QCHSC24)\\
 19-24 August, 2024\\
 Cairns Convention Centre, Cairns, Queensland, Australia\\}

\begin{document}
\maketitle

\input epsf

\def \li {\par\hskip .2in \Maroon{$\bullet$}\hskip .1 in \textBlue}
\def \lli {\par\hskip .4in \Green{$\bullet$}\hskip .1 in \textBrown}

\def \hi {\medskip \textBlack}
\def \topic #1{\par\textMaroon\centerline {#1}\ss\textBlack}
\def \nextslide{\smallskip\hrule\smallskip}
\long \def \blockcomment #1\endcomment{}

\section{Introduction}

This talk concerns how the usual $SU(3)\otimes SU(2) \otimes U(1)$
standard model fits well with a lattice formulation using Wilson
fermions and including the parity violation of the weak interaction
\cite{Creutz:2023wxu}.  The weak interactions are known violate parity
with neutrinos being left handed, and it is desirable to include this
in a mathematically well defined lattice formulation.  In addition,
while very small, in principle the weak interactions include
non-perturbative proton decay \cite{'tHooft:1976fv}.  The lattice
provides a framework to study this mechanism, rather non-intuitive in
perturbation theory.

A goal of many lattice approaches is to give a mathematical definition
of a field theory as a continuum limit.  Unfortunately we do not
realize this here because the approach relies heavily on the scalar
field associated with the Higgs mechanism.  This is used as usual to
give masses to the fermions, but also to give large masses to the
familiar fermion doublers.  Being a scalar field, the Higgs does not
satisfy asymptotic freedom, and therefore it is unclear how to take
continuum limit.  A possible direction is a composite Higgs model, but
this goes beyond the current discussion.  The $U(1)$ field of the
standard model is also not asymptotically free, but this presumably
might be sidestepped via a unified model.

The approach here is similar in spirit to
Refs. \cite{Swift:1984dg,Smit:1985nu,Shrock:1985un,Lee:1986kc}, with
the main difference being a non trivial mingling of the weak and
strong groups. For this, the appearance of an even number of fundamental weak
doublets in a single generation is essential.  Witten
\cite{Witten:1982fp} has discussed how a closed path in $SU(2)$ field
space can change the sign of the fermion determinant.  This phase
ambiguity is extensively reviewed in \cite{Poppitz:2010at} and is
closely related to the 't Hooft vertex \cite{'tHooft:1976fv}
with its connection to anomalies.

A useful ingredient is the pseudo reality of the weak group.  For
every left handed $SU(2)$ doublet of particles, their antiparticles
also form an $SU(2)$ doublet, although right handed.  Thus we have an
equal number of left and right handed fields.  Combining them into
gauge invariant singlet operators is what enables the small baryon
decay process.

On the lattice all field configurations are simply connected.  This
means that the usual continuum discussions relating anomalies to gauge
field topology must be modified.  The index theorems relating zero
modes of the Dirac operator to topology are incomplete.  We will argue
that corresponding to these modes are real eigenvalues of the Dirac
operator separated by chirality.

\begin{plain}
\section{Fields}

Consider a full generation of fermions as a set of 8 4-component
fields on lattice sites
\begin{equation}
  u^r,u^g, u^b, d^r,d^g,d^b,\nu,e^-.
  \end{equation}
Here we include the three colors for up and down quarks, denoted
$\{r,g,b\}$.  We allow all fermions to have right-handed parts,
including the neutrino, but the weak gauge fields will not couple to
these components.  In addition we include a complex Higgs doublet of
scalar fields {$H=\pmatrix{H_1 \cr H_2\cr}$} also occupying the
lattice sites.

In addition to these ``matter'' fields, we consider gauge fields
placed on the lattice bonds. Remaining as close as possible to
traditional lattice methods
\cite{Wilson:1974sk,Wilson:1975id,creutz_2023}, the bond variables
include $SU(3)$ matrices, denoted $U_{su(3}$ for the strong
interactions, $SU(2)$ matrices, denoted $U_{su(2}$ for the weak, and
finally a phase factor $U_Y$ for the hyper charge $U(1)$.  These
fields reflect the fields of the usual continuum standard model
\cite{Oerter:2006iy}.  The gauge fields self interact separately via
the standard plaquette action.  This gives three independent gauge
coupling constants.

Our single generation of fermions breaks down into two vector-like
strong $SU(3)$ triplets
\begin{equation}
u=\left(
\eqalign{u^r \cr u^g \cr u^b\cr}
\right)
\qquad 
d=\left(
\eqalign{d^r \cr d^g \cr d^b\cr}
\right)
\end{equation}
as well as 
four left handed weak $SU(2)$ doublets
\begin{equation}
r=\left(\eqalign{u^r\cr d^r\cr}\right)_L\ 
g=\left(\eqalign{u^g\cr d^g\cr}\right)_L\ 
b=\left(\eqalign{u^b\cr d^b\cr}\right)_L\ 
l=\left(\eqalign{\nu\cr e^-\cr}\right)_L.\
\end{equation}

\section{Local gauge symmetries}

The formulation preserves exact local gauge symmetries on each site.
These symmetries are independent for each group.  In particular, a
strong gauge transformation multiplies each triplet by an arbitrary $SU(3)$
element $g_{su(3)}$ on the corresponding site
\begin{equation}
\psi_{ud}\rightarrow g_{su3} \psi_{ud}.
\end{equation}
Similarly, the
weak group acts on the four left handed doublets
\begin{equation}
\psi_{rgbl}\rightarrow  \left( g_{su2} {1-\gamma_5\over 2} 
+  {1+\gamma_5\over 2}\right)\psi_{rgbl}.
\end{equation}

The gauge matrices transform as usual under the transformations
associated with the ends of the corresponding bonds
\begin{equation}
  \matrix{
U_{su3}^{ij}\rightarrow g_{su3}^i\  U_{su3}^{ij}\ {g^j_{su3}}^\dagger\cr
U_{su2}^{ij}\rightarrow g_{su2}^i\  U_{su2}^{ij}\ {g^jr_{su2}}^\dagger.\cr
}
\end{equation}
Finally, the weak $SU(2)$ gauge group also acts on Higgs fields
\begin{equation}
  H=\pmatrix{H_1\cr H_2\cr}
  \rightarrow g_{su2}H.
  \end{equation}

The pseudo-reality of $SU(2)$ says that each 
element is similar to its complex conjugate
\begin{equation}
  g^*=\tau_2 g \tau_2.
\end{equation}
This immediately implies that another combination of the Higgs field
\begin{equation}
  H^\prime\equiv\tau_2 H^* \tau_2 =
\pmatrix{-H_2^*\cr H_1^*\cr
}\end{equation}
transforms under gauge transformations equivalently to $H$
\begin{equation}
  H^\prime
  \rightarrow g_{su2}H^\prime.
\end{equation}

Finally the hyper-charge introduces phases $g_Y$ on the fields
proportional the corresponding hyper-charge $\psi\rightarrow g_Y
\psi$.  These values take the same values they take in conventional
continuum discussions.  For the fundamental fields
\begin{equation}
u^r,u^g, u^b, d^r,d^g,d^b,\nu,e^-
\end{equation}
the corresponding hyper-charge values are
\begin{equation}
Y_L=(1/3,1/3,1/3,1/3,1/3,1/3,-1,-1)= 2Q\pm 1
\end{equation}
for the left hand parts and
\begin{equation}
Y_R=(4/3,4/3,4/3,-2/3,-2/3,-2/3,0,-2)= 2Q
\end{equation}
for the right hand parts.
The Higgs fields also have hyper-charge, with
$Y_H=1$,  $Y_{H^\prime}=-1$.
Finally the gauge fields are all neutral under hyper-charge.


An important observation is that the gauge symmetries under $SU(3)$,
$SU(2)$ and $U(1)$ groups all commute.  In particular the weak group
$SU(2)$ doesn't change $SU(3)$ colors, the strong group doesn't break
weak chirality, and hyper-charge is constant on each multiplet,
whether strong or weak.

\section{Composite fields}

From each fermion doublet we can form an $SU(2)$ gauge singlet with the
Higgs field in two different ways $(H, \psi_L)$ and
$({H^\prime},\psi_L)$.  More precisely, the singlets are defined
\begin{equation}
  (H, \psi_L)\equiv H_1^*\psi_1+H_2^*\psi_2
\end{equation}
and
\begin{equation}
  (H^\prime, \psi_L)\equiv -H_2 \psi_1+H_1 \psi_2.
\end{equation}
There are thus two SU(2) invariant combinations per doublet.  These
combinations represent the physical left handed particles, which might
be interpreted as ``composite.''  To make closer contact with
continuum discussions, one can divide out the Higgs ``vacuum
expectation'' $v=|H|$ and obtain physical combinations
\begin{equation}
\matrix{
&e_L&=&(H,l)/v             && Q=(Y_l-Y_H)/2=-1\cr\cr
&\nu_L&=&({H^\prime},l)/v   && Q=(Y_l+Y_H)/2=0\cr\cr
&{u_{rgb}}_L&=&(H,rgb)/v && Q=(Y_{rgb}-Y_H)/2=2/3\cr\cr
&{d_{rgb}}_L&=&({H^\prime},rgb)/v  && Q=(Y_{rgb}+Y_H)/2=-1/3.\cr
}
\end{equation}
This extraction is equivalent to the perturbative ``unitary'' gauge.
As the Higgs field spins around in gauge space, the fermion fields
follow along with them.

\section{Masses and doublers}

As in the usual continuum discussions, the fermions are given masses
via the Higgs mechanism
\cite{Higgs:1964pj,Weinberg:1967tq,Salam:1968rm,Englert:1964et,Guralnik:1964eu}.
For this lattice construction we use the above $SU(2)$ invariant
combinations
\begin{equation}
\chi_L={1\over v}
\pmatrix{(H, \psi_L)\cr {(H^\prime}, \psi_L)\cr}
\end{equation}
to mix with the left handed fermions and give
 gauge singlet mass terms
\begin{equation}
  \overline\psi_R M  \chi_L + h.c.
\end{equation}.

Once we have given the fermions their masses, we can use a similar
mechanism to drive the doubler masses to large values, in the
``cutoff'' region.  This uses the conventional Wilson term
\cite{Wilson:1975id} modified to use these ``physical'' fields
\begin{equation}
 {\overline\psi_R}_{i+e_\mu}(1+\gamma_\mu){\chi_L}_i/2
  +{\overline\psi_R}_{i}(1-\gamma_\mu){\chi_L}_{i+e_\mu}/2
  + h.c.
  \end{equation}
This term mimics the ``formally irrelevant'' operator that is only
important in the large momentum doubler region.  As in usual
discussions of Wilson fermions, this terms breaks chiral symmetry and
requires an additive mass tuning.  In particular, the approach offers
no explanation of why neutrinos are so light.

\section{Physical gauge fields}

The bare gauge fields are located on the lattice bonds, which we label
by their endpoints $ij$.  As with the fermion fields, we can combine
the bond variables with the Higgs field to create gauge invariant
operators for the physical weak bosons.  For example, the combination
\begin{equation}
  W^+=(H^\prime_i, {U_{su2}}_{ij} H_j)
\end{equation}
has electric charge one and represents a field creating the $W^+$
meson.  Correspondingly,
\begin{equation}
  W^-=(H_i, {U_{su2}}_{ij} H^\prime_j)
\end{equation}
creates the negative $W$ boson.

Following this general scheme, there are two neutral weak operators
available on the bonds
\begin{equation}
  \eqalign{
    &(H^\prime_i, {U_{su2}}_{ij} H^\prime_j)\cr
    &(H_i, {U_{su2}}_{ij} H_j).\cr
  }
\end{equation}  
These in general will mix, with one combination representing the $Z$
and the other the hopping parameter for the
physical Higgs boson.

\section{Anomalies and Dirac eigenvalues}

This basically completes the model, but it is instructive to consider
how the usual quantum anomalies come into play.  With dimensional
regularization these effects appear via the fermionic measure not
being chirally symmetric \cite{Fujikawa:1979ay}.  With Wilson
fermions, the anomalies are moved into the behavior of heavy doubler
states.  This is similar in spirit to discussions of anomalies
with Pauli-Villars regulation
\cite{Pauli:1949zm,Adler:1969er,Bell:1969ts}, with additional heavy
states added near the cutoff.

In continuum discussions anomalies are frequently tied to topology in the
gauge fields and the index theorem \cite{Atiyah:1963zz,Atiyah:1971rm}
relating to zero modes in the Dirac operator.  On the lattice the
space of allowed configurations is simply connected and does not
support separate topological sectors absent some sort of smoothing
condition \cite{Luscher:1981zq}.  However such restrictions destroy
reflection positivity \cite{Creutz:2004ir} and will interfere with any
Hamiltonian formulation.

It is interesting to contrast this picture with the overlap approach
of Neuberger \cite{Neuberger:1997fp,Narayanan:1992wx,Narayanan:1993sk}
where one projects the relevant eigenvalues onto exact zero modes.
This has been successful to all orders in perturbation theory
\cite{Luscher:2000zd}.  It does, however, eliminate ultra-locality of
the Dirac operator \cite{Horvath:1998cm,Horvath:1999bk}.  In addition
the projection process encounters singularities as one transits
between topological sectors \cite{Creutz:2002qa}.  In the current
approach, the Dirac operator remains local while robust zero modes are
lost.

Our Dirac operator has been constructed to satisfy
gamma 5 hermeticity
\begin{equation}
\gamma_5 D \gamma_5= D^\dagger.
\end{equation}
Indeed most lattice fermion prescriptions, with the exception of
twisted mass \cite{Frezzotti:2000nk}, satisfy this.  This implies that
all eigenvalues of $D$ are either in complex conjugate pairs or
real. As a reminder, in Fig. \ref{eigen1} we sketch the eigenvalue
spectrum for free Wilson fermions.  When non-trivial gauge fields are
turned on, these eigenvalues move around.  If the gauge fields are
sufficiently smooth, the index theorem does apply and modes of
non-trivial chirality are well known.  However, since the space of
lattice fields is simply connected, there must exist a continuous path
connecting a configuration without such chiral states to one with
them.  As discussed in Ref. \cite{Creutz:2002qa} in terms of the
eigenvalues of $D$, a complex conjugate pair of eigenvalues can join
on the real axis and split apart as two real eigenvalues.  One of
these can move to have a small real part while the other moves off to
the doubler region, as sketched in Fig. \ref{eigenflow}.
Ref. \cite{Creutz:2010ec} demonstrated such a path does not need to
pass through a barrier of of large action, although it does require
local fields to violate the smoothness condition of
\cite{Luscher:1981zq}.

\begin{figure}
 \centerline{ \includegraphics[width=.4\hsize]{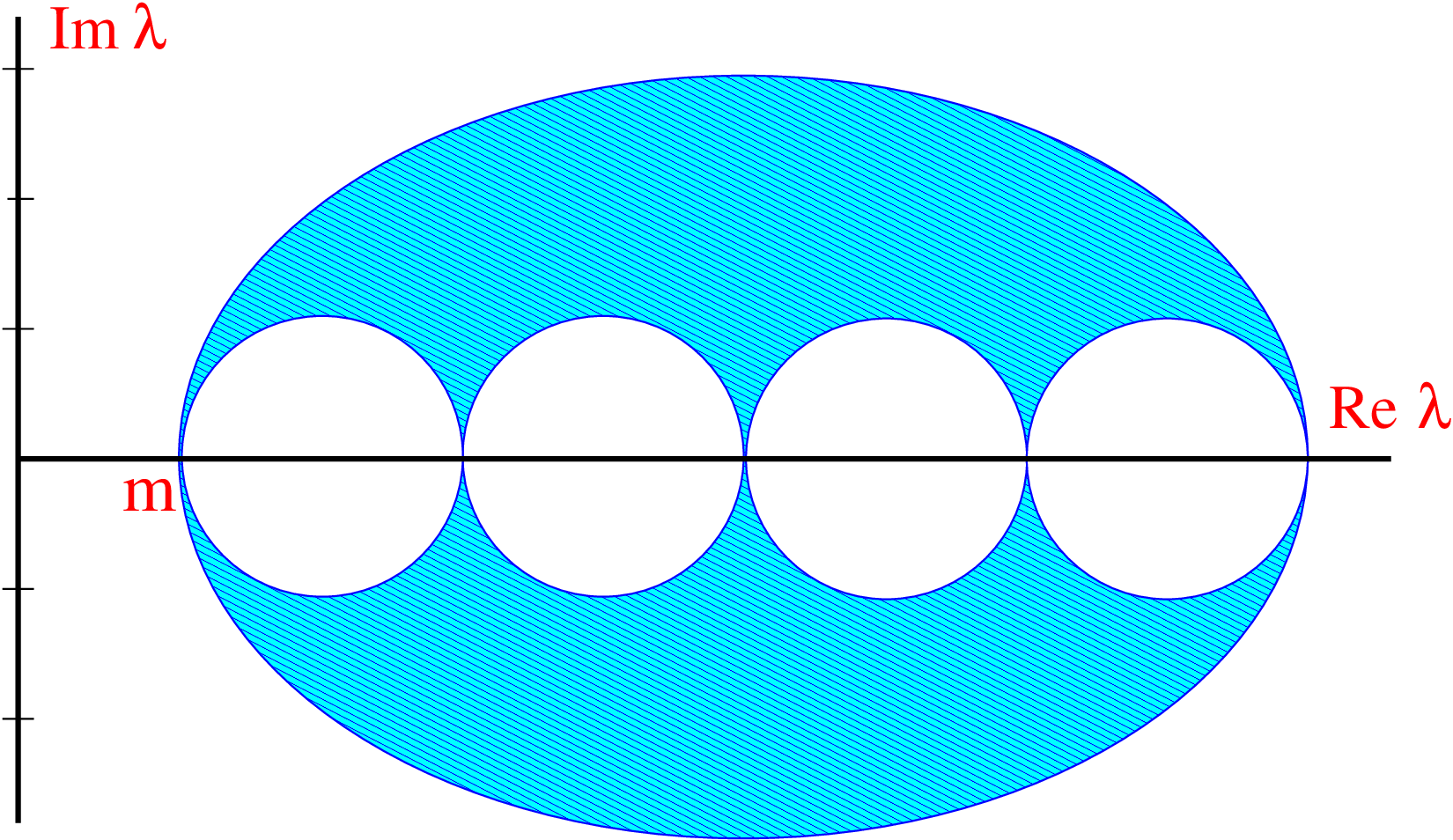}}
 \caption{ The spectrum of free Wilson fermions.  The Wilson term
   moves the doubler eigenvalues to large values in the cutoff region.
 }
\label{eigen1}
\end{figure}

\begin{figure}
 \centerline{ \includegraphics[width=.4\hsize]{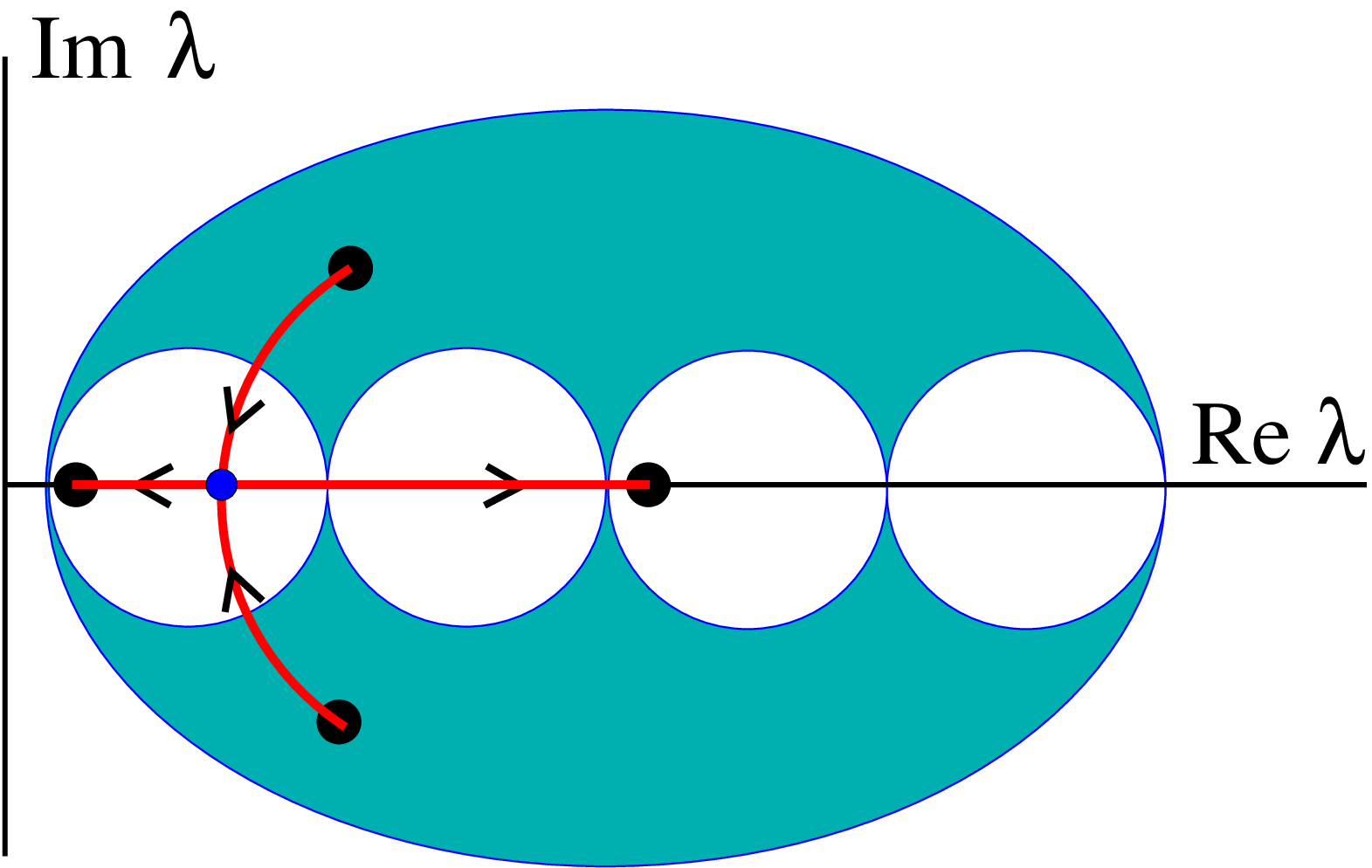}}
 \caption{ As we travel through configuration space, two complex
   conjugate Dirac modes can join on the real axis and become a pair
   of real eigenvalues of opposite chirality.  }
\label{eigenflow}
\end{figure}

Restricted to the space spanned by real eigenmodes, $\gamma_5$ commutes
with the Dirac operator.  On this space $\gamma_5$ and $D$ can be
simultaneously diagonalized.  Therefore all the real eigenvalues can
be sorted by chirality.  More precisely, if we have some isolated mode
satisfying
\begin{equation} D \psi=\lambda \psi
\end{equation}
with lambda real, this mode also satisfies
\begin{equation}
  \gamma_5 \psi=\pm 1.
  \end{equation}

The concept of ``topology'' from the continuum field theory is
replaced by an excess of small eigenvalues of one winding.  On the
lattice, these chiral real modes remain robust under small
deformations of the fields.  For smooth fields with a differentiable
continuum limit, this reduces to the index theorem.

\section {The 't Hooft process}

Small eigenvalues of $D$ are suppressed in the partition function
\begin{equation}
Z=\int (dA)(d\overline \psi d\psi)
\ e^{-S_g+\overline\psi D \psi}
=\int (dA)\ e^{-S_g(A)}\ \prod \lambda_i.
\end{equation}
This would suggest that any approximate zero modes would become
irrelevant and small eigenvalues are of little importance.  This naive
view was shown to be incorrect by 't Hooft \cite{'tHooft:1976fv}.  The
point being that certain observables can enhance the small modes.

To see this, introduce sources $\eta$ and $\overline \eta$,
differentiation with respect to which generates Green's functions
\begin{equation}
Z(\eta,\overline\eta)=\int (dA)\ (d\overline \psi d\psi)\ 
e^{-S_g+\overline\psi D \psi +\overline\psi \eta+
\overline\eta\psi}.
\end{equation}
Completing the square gives
\begin{equation}
Z=\int (dA)\ 
e^{-S_g{+\overline\eta {D^{-1}} \eta}/4}\ 
\prod \lambda_i. 
\end{equation}
The factor of $D^{-1}$ can bring in the inverse of small eigenvalues
and cancel any suppression.


How this works to give the chiral anomaly in the strong interactions
is well understood.  The needed observables couple left and right
fermions $\langle\psi_R D^{-1} \psi_L\rangle\ne 0$.  The effective
vertex applies to all strong triplets and we get a coupling that give
a mass to the flavor singlet eta prime meson, as sketched in
Fig. \ref{fourpoint}

\begin{figure}
 \centerline{ \includegraphics[width=.4\hsize]{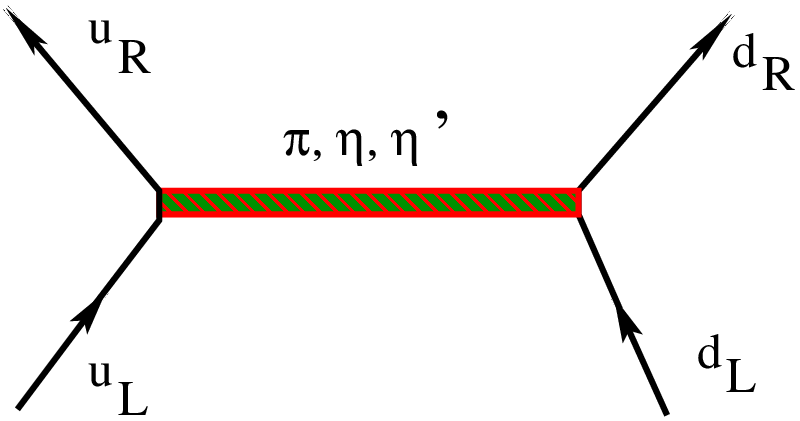}}
 \caption{Small chiral eigenvalues involving the strong group generate
   an effective interaction that contributes to the mass of the flavor
   singlet eta prime meson.  Figure taken from
   Ref. \cite{Creutz:2010ts}}
\label{fourpoint}
\end{figure}

\section{Weak interactions}

Because the electroweak coupling is so small, the effects of the
anomaly for the weak interactions are highly suppressed and
essentially unobservable.  However the consequences include the
non-conservation of baryon and lepton number and are rather
non-intuitive.

We start with our four left handed doublets. For each of them, their
antiparticles are also a doublet, although right handed
\begin{equation}
  \psi^c_R=\tau_2\gamma_2\psi_L^*.
  \end{equation}

For our observable, we want to combine these fields into a Lorentz
invariant and gauge invariant combination.  We start by pairing each
doublet with a second conjugate doublet

\begin{equation}
 \overline \psi_{iR}^c D^{-1} \psi_{jL}
\end{equation}
where the indices run over the four doublets $i,j \in \{r,g,b,l\}$.
The $D^{-1}$ removes zero mode suppression.  Antisymmetrization over
the doublets restores strong gauge invariance
\begin{equation}
\epsilon_{ijkl}
\ \langle
\overline \psi_i^c D^{-1} \psi_j
\ \overline \psi_k^c D^{-1} \psi_l\rangle
\ \ne 0.  
\end{equation}

This effective vertex changes baryon number and lepton number each by
1 but preserves $B-L$.  In a Hamiltonian approach this involves modes
crossing into and out of the Dirac sea
\cite{Creutz:1993vv,Creutz:2001wp}.  In the process Fermion number
changes by 2.  This is consistent with $SU(2)$ since the group is
pseudo-real.  It is also consistent with $SU(3)$ since $\overline 3
\in 3\otimes 3$ and we are taking two flavors to one anti-flavor.
More physically, the process involves an ``effective'' neutron
anti-neutrino and proton positron mixing
\begin{equation}
  \pmatrix{n\cr  p} \Longleftrightarrow \pmatrix {\overline\nu \cr e^+}.
\end{equation}
Through this mixing the process
\begin{equation}
  p\rightarrow e^+ + \pi
\end{equation}
is allowed. Of course it is st a very small rate of order
$O(e^{-c/\alpha})$ with $c$ involving the action of the chiral
eigenvalue.  A version of this vertex appears in a proposed domain
wall approach to the weak interactions
\cite{Creutz:1996xc,Creutz:1997fv,Creutz:2018dgh}.

\section{Summary}

We have discussed how a single generation of the standard model fits
nicely onto a Wilson lattice.  In the construction the $SU(3)\otimes
SU(2)\otimes U(1)$ gauge symmetries commute and remain exact.  It
requires including a full full generation of quarks and leptons.  We
find a mechanism for baryon and lepton number violation, while the
combination $B-L$ is preserved.  The model has essentially the same
parameters as in continuum discussion: fermion masses, gauge
couplings, and the Higgs potential.

The main remaining issue concerns asymptotic freedom, absent both in
electromagnetism and the Higgs quartic self coupling.  This leaves an
obstacle towards defining the theory as a continuum limit.  For
electromagnetism this suggests a possible unification with further
gauge fields at high energies.  For the Higgs it hints at composite
models or possibly involving gravity at the highest energies
\cite{Shaposhnikov:2009pv,Alekhin:2012py}.

\end{plain}

\section*{Acknowledgment}
This manuscript has been authored under contract number
DE-AC02-98CH10886 with the U.S.~Department of Energy.  Accordingly,
the U.S. Government retains a non-exclusive, royalty-free license to
publish or reproduce the published form of this contribution, or allow
others to do so, for U.S.~Government purposes.




\end{document}